\documentclass{iaus}
\usepackage{graphicx}

\title[Spectra of Nearby Galaxies] 
{Spectra of Nearby Galaxies Measured with a New Very Broadband Receiver}

\author[Narayanan et al.]   
{Gopal Narayanan$^{1,2}$, Ronald L. Snell$^1$, Neal R. Erickson$^1$, Aeree
  Chung$^1$, Mark H. Heyer$^1$, Min Yun$^1$, \and William M. Irvine$^{1,3}$}

\affiliation{$^1$Astronomy Department, University of Massachusetts, Amherst, MA 01003 USA
\\[\affilskip]$^2$email: {\tt gopal@astro.umass.edu} 
$^3$The Goddard Center for Astrobiology}

\pubyear{2008}
\volume{251}  
\pagerange{1--4}
\jname{Organic Matter in Space}
\editors{S.K. Wok \& S. Sandford, eds.}
\begin{document}

\maketitle

\begin{abstract}
Three-millimeter-wavelength spectra of a number of nearby galaxies
have been obtained at the Five College Radio Astronomy Observatory
(FCRAO) using a new, very broadband receiver. This instrument, which
we call the Redshift Search Receiver, has an instantaneous bandwidth
of 36 GHz and operates from 74 to 110.5 GHz. The receiver has been built
at UMass/FCRAO to be part of the initial instrumentation for the Large
Millimeter Telescope (LMT) and is intended primarily for determination
of the redshift of distant, dust-obscured galaxies. It is being tested
on the FCRAO 14~m by measuring the 3~mm spectra of a number of nearby
galaxies. There are interesting differences in the chemistry of these
galaxies.  
\end{abstract}

\firstsection 
\section{Introduction}

The Large Millimeter Telescope (LMT) is a 50-meter diameter
millimeter-wavelength single-dish telescope being built jointly by the
University of Massachusetts, Amherst in the USA and the Instituto
Nacional de Astrof\'isica, \'Optica y Electr\'onica in Mexico
(\cite{alfonso2006}). The telescope uses recent advances in structural
design and active control of surface elements, and aims to reach an
overall effective surface accuracy of $\sim 70$ $\mu$m and an ultimate
pointing accuracy of better than 1$^{\prime\prime}$. The LMT is sited
at 4600~m elevation at a latitude of 19$^\circ$~N in the Mexican state
of Puebla and offers good sky coverage of both hemispheres. The
normally low humidity will allow operation at frequencies as high as
345 GHz. Telescope construction is well advanced. Three of the planned
five rings of surface panels are in place. The initial complement of
instruments will include SEQUOIA, a 32 element heterodyne focal plane
array for 3mm that is currently in use at FCRAO; AzTEC, a large
format, focal plane bolometer array that has had successful runs on
the JCMT and ASTE; a dual-polarization receiver for the 1mm band; and
a unique wide band receiver and spectrometer, the Redshift Search
Receiver (RSR), the instrument utilized in the present paper.

The RSR (\cite{erickson2007}) utilizes very wideband indium phosphide
MMIC amplifiers operated at 20 K and has two dual polarization beams
(thus a total of 4 receivers). Heterodyne receivers require a fast
beam switch to produce flat spectral baselines over such a wide
bandwidth, and the RSR uses a novel polarization switch operating at
1~kHz (\cite{erickson2007a}). This is the first wide band, low loss
electrical switch operating at a wavelength as short as 3~mm. The fast
polarization switch is followed by a broadband orthomode transducer
(OMT) that splits the polarizations into two independent
receivers. One beam of each polarization of the RSR is always on
source.

The principal motivation for the construction of the RSR is to measure
the spectra, and particularly the redshift {\tt z}, of very distant
galaxies. Galaxies are believed to form in the very early universe
with the first episode of star formation $\sim 10^9$ years after the
Big Bang, corresponding to ({\tt z} $\gtrsim 6$).  Understanding the
formation process requires a catalog of a significant population out
to {\tt z} $\sim 10$ ($5\times10^8$ yrs). Although these objects are
relatively easy to detect in continuum, their distance and age are not
easy to determine.

A few of these galaxies have redshifts measured in the visible, but
most have no visible counterpart. The strongest emission lines from
galaxies at roughly mm-wavelengths are the rotational transitions of
CO, and the RSR bandwidth is large enough that there is a very high
probability that one line of CO will fall in the observing band for
redshifts {\tt z} $>1$, and that two CO rotational lines will be in
the band for {\tt z} $>3.2$. When two lines of the CO rotational
ladder are detected within the RSR band, the redshift is uniquely
determined.

Since the LMT is not yet complete (we are hoping for initial 3~mm
commissioning during 2008), the RSR was installed on the FCRAO 14~m
telescope during spring 2007. The receiver frontend worked very well
with the spectrometer to give very flat baselines. CO emission from 22
ULIRGs at moderate redshift was detected (\cite{chung2008}), and
broadband 3~mm spectra of many nearby galaxies were obtained. The
latter are presented here.

\section{Observations}
\label{observations}

Each pixel of the RSR is sent to 6 different backend cards, each of
which can process $\sim 6.5$ GHz of bandwidth. Twenty-four backend
cards are required to process all four pixels over the 75--111 GHz
bandwidth. While all four frontend pixels were available for the
Spring 2007 run, only 4 out of the required 24 sections of the RSR
backend spectrometer had been fabricated at that time, and were ready
to use. For the nearby galaxy work, these four spectrometer cards were
hooked up to the best pixel of the frontend to cover first a bandwidth
range of 85 -- 111 GHz, and then in separate observations a frequency
range of 75 -- 92 GHz. The two sets of observations are combined and
averaged together (using special-purpose software written specifically
for the RSR) to produce the spectra presented here.


\section{Results and Discussion}

\begin{figure}[!htb]
\scalebox{0.57}{\includegraphics[angle=-90]{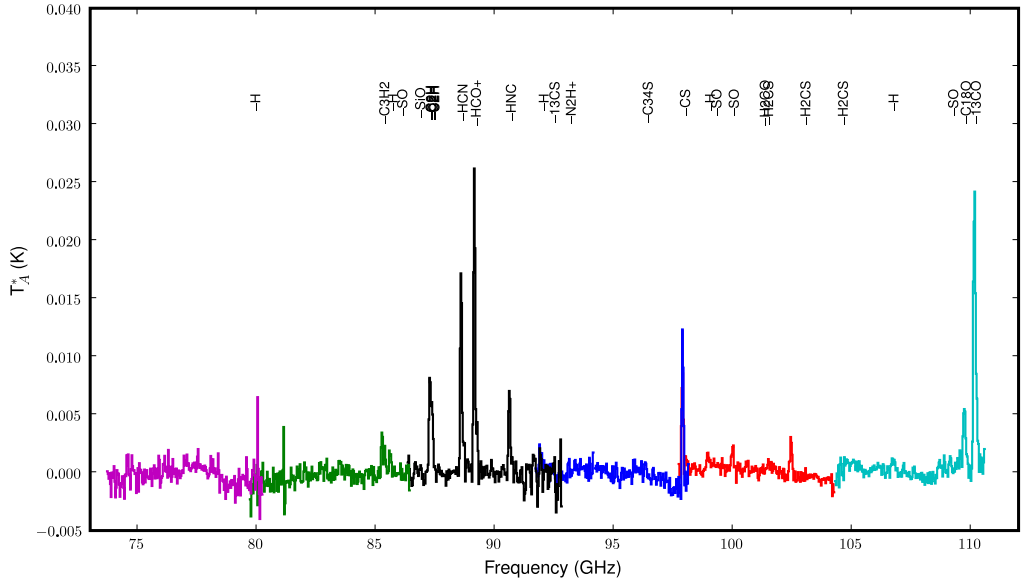}}
 \caption{Full 3~mm band spectrum of M82 obtained with the redshift
   search receiver. The different colors represent distinct
   spectrometer boards that were used in the measurment. The names of
   important molecular lines typically seen in the ISM are denoted at
   the appropriate frequencies in the plot. M82 has strong continuum
   emission, and that accounts for the non-flat baselines (compare
   with Fig~\ref{ic342}).}
   \label{m82}
\end{figure}

\begin{figure}[!htb]
\scalebox{0.57}{\includegraphics[angle=-90]{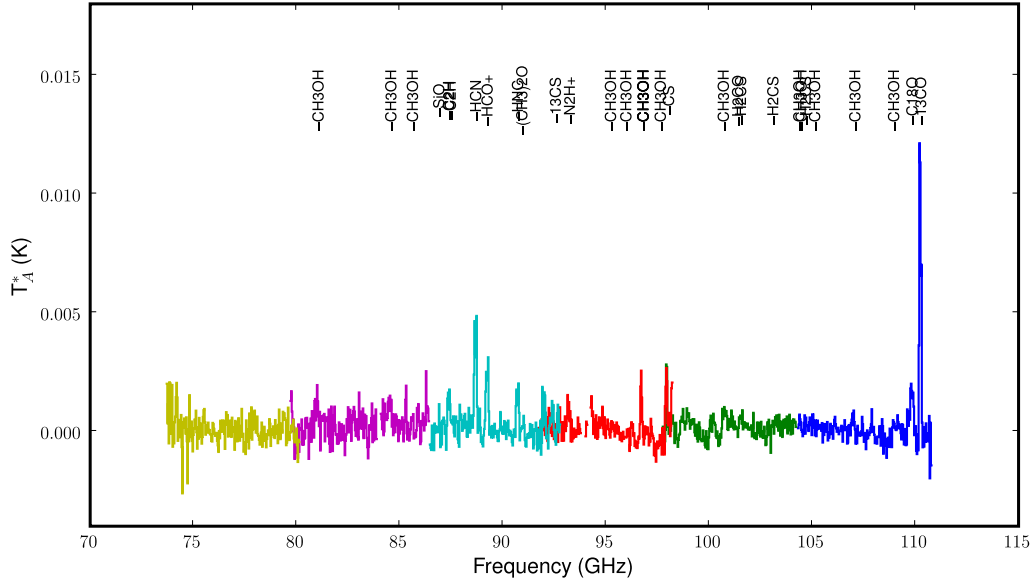}}
 \caption{Full 3~mm band spectrum of IC342 (see Fig~\ref{m82}).}
   \label{ic342}
\end{figure}

In Figures~\ref{m82} and \ref{ic342}, we show the full 3~mm band
spectra towards two galaxies, M82 and IC342. The emission lines from a
number of molecular species are detected in the observed galaxies (see
Table~\ref{tab1}). Note that the J=1-0 transition of $^{12}$CO,
normally the strongest 3~mm emission line from galaxies, is not
included in the observed bandwidth, but would move into the band for
moderate values of the redshift {\tt z}. As has been found by previous
observers (e.g., \cite{aalto2006, aalto2007, martin2008}), the ratios
of the integrated intensities of such important chemical tracers as
HCN, HNC, HCO$^+$, N$_2$H$^+$, CS and $^{13}$CO vary among the
galaxies. It has been proposed that such variations may be due to
differences in the relative importance of photon-dominated regions
(PDRs); X-ray dominated regions (XDRs), perhaps in the vicinity of
AGNs; shocked regions; warm, dense molecular clouds; and infrared
pumping for HCN and HNC (e.g., \cite{aalto2008, imanishi2007,
  perez2007, turner2008}). Because of the rather large beam-size of
the FCRAO 14~m telescope (from 45 -- 65 arcsec), it is difficult to
definitely separate these effects for our galaxies; this will be much
more clear cut with the LMT, whose corresponding beam size will be in
the range 12 -- 18 arcsec. Nonetheless, we note that our results for
the ratios of the integrated intensities HCO$^+$/HCN, HNC/HCO$^+$, and
HNC/HCN (see Table~\ref{tab1}a) agree well for M82, NGC 253 and Arp
220 with those reported by \cite{baan2008} (observations with the IRAM
30m and the SEST 15m), although those latter authors give somewhat
higher values for the HNC/HCN ratio in NGC 253.

\begin{table}
  \begin{center}
\begin{minipage}{17pc} 
   {\scriptsize
  \begin{tabular}{|l|l|l|l|l|}\hline
{\bf Ratio} & {\bf NGC~253} & {\bf M82} & {\bf IC342} & {\bf Arp220}\\
\hline
{\bf HNC/HCN} & 0.51 & 0.43 & 0.37 & 0.81\\
&(0.03) & (0.02) & (0.06) & (0.11)\\
{\bf HCO$^+$/HCN} & 0.84 & 1.47 & 0.64 & 0.50 \\
&(0.03) & (0.04)& (0.07) & (0.11)\\
{\bf HCN/$^{13}$CO} & 0.92 &0.87 & 0.55 & 1.8\\
&(0.03) & (0.03) & (0.03) & (0.6)\\
{\bf CS/HCN} & 0.43 & 0.58 & 0.33 & 0.55\\
& (0.02) & (0.03) & (0.05) & (0.14) \\
{\bf N$_2$H$^+$/HCN} & 0.06 & $< 0.02$ & 0.24 & 0.45\\
& (0.02) && (0.07)&(0.08)\\
\hline
  \end{tabular}
}
\end{minipage}
\begin{minipage}{12pc} 
  {\scriptsize
\begin{tabular}{|l|l|l|}
\hline
{\bf Line} & {\bf Frequency (GHz)} & {\bf Transition}\\\hline
C$^{18}$O & 109.8 & 1-0\\
HC$_3$N & 109.1 & 12-11\\
CH$_3$C$_2$H & 102.5 & 6(n)-5(n)\\
HC$_3$N & 100.1 & 11-10\\
SO & 99.3 & 3(2)-2(1)\\
CH$_3$OH & 96.7 & 2(0,2)-1(0,1)\\
&&A+, \& blend\\
C$_2$H & 87.3 & 1-0, 3/2-1/2 \\
CH$_3$C$_2$H & 85.5 & 5(n)-4(n)\\
C$_3$H$_2$ & 85.3 & 2(1,2)-1(0,1)\\
CH$_3$OH & 84.5 & 5(-1,5)-4(0,4) E\\
HC$_3$N & 81.9 & 9-8 \\
CH$_3$OH & 81.0 & 7(2,6)-8(1,7) A-\\
\hline
  \end{tabular}
}
\end{minipage}
\caption{(a) Molecular Line Ratios in 4 Galaxies (J=1-0 except CS,
  J=2-1). (b) Other detected lines.}
  \label{tab1}
\end{center}
\end{table}

\section{Summary}

Spectra in the 3-mm wavelength band covering 75 to 111 GHz have been
observed for about 10 galaxies using a new, very broadband receiver
(RSR) with the FCRAO 14m radio telescope. Interesting differences in
line ratios are found, consistent with previous observations. When the
RSR is mounted on the LMT in Mexico, surveys of the chemistry of
external galaxies will be possible. We note in this connection the
suggestions that the variety of environments near the center of the
Milky Way may serve as templates for unraveling the processes in
external galaxies (\cite{martin2008a, jones2008}).

{\bf Acknowledgements:} We are grateful for partial support of this
research by the NSF (AST-0096854 and AST-0704966) and by NASA GSFC
Cooperative Agreement NNG04G155A; and to Mike Brewer, Ron Grosslein,
Kamal Souccar, Don Lydon and Gary Wallace for assistance during the
engineering commissioning of the RSR at the FCRAO 14m telescope.

\end{document}